# LLM-based Vulnerability Discovery in Business Process Documentation

B.Falchuk, H.Garg, E.Panagos, S.Zohar

[1] Peraton Labs, Basking Ridge, USA
{bfalchuk|himanshu.garg|epanagos}@peratonlabs.com, sioan.zohar@peraton.com[2]

**Abstract.** Just like software and hardware, business processes are susceptible to vulnerabilities that can lead to product quality issues, delays, and increased costs. Business process vulnerabilities can arise from a variety of sources, including conflicting requirements, ambiguous documentation, invalid measurement specifications, omission of quality checks, or implementations that differ from specifications. MIRABELLE is a system that identifies and characterizes business logic (BL) vulnerabilities from available business process representations, including ISO 9000/9001 documentation, user guides, work instructions, and process execution logs. MIRABELLE leverages recent advances in AI/ML to process available business process documentation and generate attributed graph representations of the business logic that can be processed using both graph and formal logic approaches for identifying potential vulnerabilities. However, extracting the business logic (e.g., operation execution sequences, decisions, input/output resources) from mostly natural language artifacts is challenging due to the required domain expertise, inherent process complexity, and the sometimes very large volumes of information. This paper focuses on our experimentation with Large Language Models (LLMs) and their role within MIRABELLE. We report on the performance of several LLMs across vital stages of vulnerability detection, from grammatical and technical error-flagging in short phrasings, to complete process structure recovery and extraction.

## 1 Introduction

MIRABELLE (Multimodal Ingestion, Representation, and Analytics of Business and Enterprise Logic for Limiting Exploits) identifies and characterizes business logic (BL) faults and vulnerabilities from extracted business process representations.

[1] This research was developed with funding from the Defense Advanced Research Projects Agency (DARPA). The views, opinions and/or findings expressed are those of the author and should not be interpreted as representing the official views or policies of the Department of Defense or the U.S. Government. Distribution Statement "A" (Approved for Public Release, Distribution Unlimited).

[2] Author 1 and 2 contributed equally to this work.

MIRABELLE, shown in Figure 1, ingests business logic (BL) system artifacts, such as human-entered code, ISO 9000/9001 documentation, user guides, and system information, and generates BL Model Representations using a combination of techniques, including a novel application of large language models (LLMs). These representations – which may be incomplete - are processed by a BL Model Synthesis phase which creates an aligned BL model via entity alignment, mapping it to logic formulas, and composing these formulas using logical operators. The resulting formal model is processed by BL Flaw Analyzer for identifying flaws associated with control flows, input/output data artifacts, and sub-process interactions. Extracted BL workflows and identified flaws are presented via easy-to-use human interfaces and in common syntax such as JSON or Business Process Modeling and Notation (BPMN) notation [21].

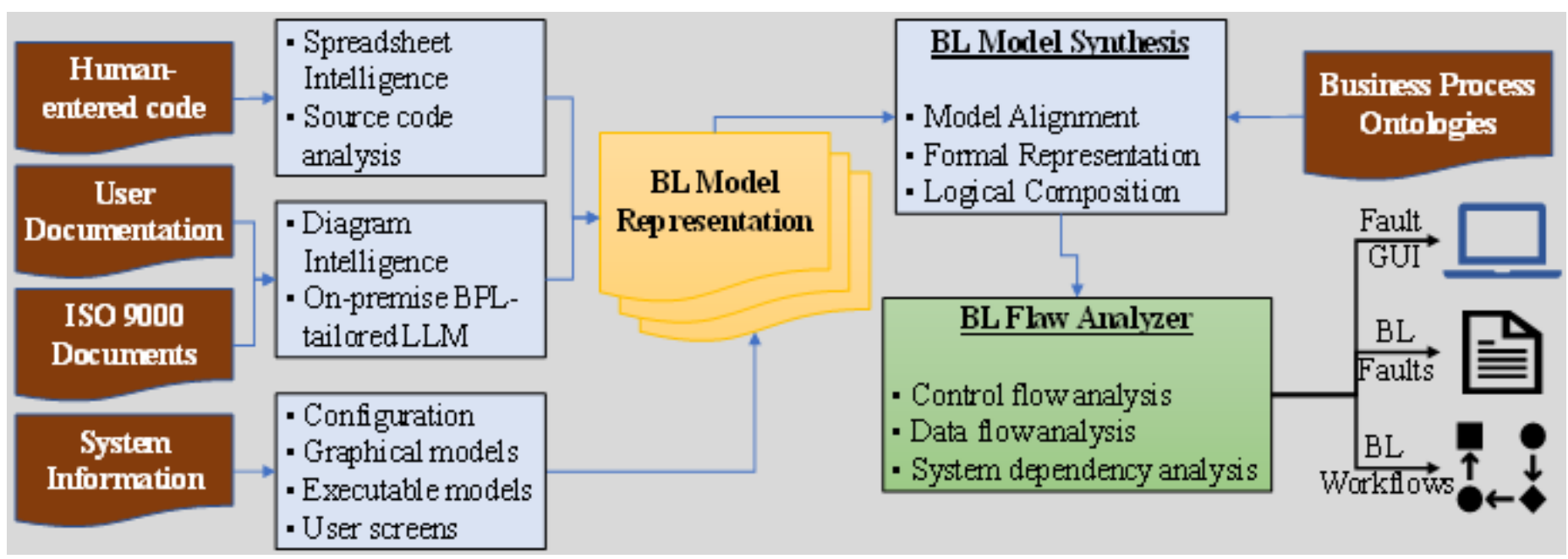

**Fig.1**. MIRABELE high-level architecture.

By reasoning about business logic faults using formal representations, MIRABELLE supports a wide range of Enterprise Resource Planning (ERP) and Manufacture Execution Systems (MES) applications across different business domains. Although the business process intelligence and re-engineering community has developed several tools (e.g., SAP Signavio [11], Celonis [10], ProM [12], Camunda [13]) for analyzing business processes, their focus is on identifying performance bottlenecks and reengineering, and they require a robust connection to the operational BL system instance. In contrast, MIRABELLE focuses on adversarial process intelligence to identify and characterize business logic flaws with limited visibility and access to the BL platform. Adversarial process intelligence includes BL model ingestion and representation and vulnerability characterization.

Other approaches for BL model ingestion that use natural language processing (NLP) techniques (e.g., [14][15][16]) are limited in handling complex processes in Defense Industrial Base (DIB) systems. These techniques use part of speech (POS) semantic tagging to generate parse trees that are then used for mapping nodes and actors, activities, and execution flows. Typically, semantic tagging uses a lexicon for tagging words with corresponding BPMN entities. For example, a sentence that includes "while" or "meantime" may be tagged as representing a parallel gateway. Although NLP-based approaches work well when the process description in natural text is well structured and uses specific terms to denote activity flows, they fail to generalize and address the free text descriptions that may be included in ISO compliance-related doc-uments and user guides without substantial effort in both updating and maintaining lexicons and reasoning about sub-process interactions.

In this paper, we describe LLM-based techniques used by MIRABELLE for extracting business process logic representations from available documentation and identifying potential business logic fault due to incorrect or contradicting instructions used by humans during the execution of specific process steps (e.g., assembling a turbofan engine). Our document processing pipeline integrates structural preprocessing, LLM-based extraction, and workflow synthesis for generating accurate, machine-interpretable business logic representations without manual intervention. Initial experimentation results across several LLM models and hardware resources show that non-quantized models deliver the highest-fidelity business logic outputs, while quantized models offer strong extraction performance and superior throughput. These findings validate the feasibility of large-scale automated ERP documentation processing and identify opportunities to enhance structured reasoning within quantized, CUI-restricted environments.

## 2 Related Work

Several academic research efforts have focused on generating business process models from natural language descriptions (e.g., [14][15][16]). Such approaches depend on natural language processing techniques (NLP), including part of speech (POS) tagging, syntax tree construction, and custom lexicons for mapping certain words and phrases to process activities and flows. For well-structured and succinct descriptions of business process activities and ordering constraints in natural text, NLP-based approaches can recover the business process model, and even generate its BPMN representation, with high accuracy (e.g., [16] claims 81% accuracy). However, these approaches are limited to only recovering relatively small (in terms of activities and execution order) business process representations. In contrast, MIRABELLE scales to large business process models by exploiting the power of large language models (LLMs) and recent advances in running such models on modest hardware.

Existing process intelligence and workflow synthesis solutions can only infer partial process models that miss certain details. For example, an approach based on process mining can only infer transactions and states that have been executed during the collection phase of the system traces. This might result in missing important but infrequently executed processes (e.g., payroll, which is only run once a month). At the same time, process intelligence might also infer transactions or states that are not reachable in the actual systems. This might, for example, happen when process models are inferred from user documentation and different terms for the same concepts are mapped to different states (e.g., some documents might use “account number” while other refer to “cost code” or “cost center”). To address this problem of missing or superfluous data/transactions/states in the synthesized workflow models, MIRABELLE uses a novel approach for checking the consistency of extracted workflow models using a combination of techniques and reference models that are part of MIRABELLE’s process ontology.

Existing research works that reason about control and data flow errors in business processes are limited in several ways. In particular, existing approaches to data flow error detection in business processes focus on missing (data element accessed before initialization), redundant (produced data element by an activity not used by follow-up activities), and conflicting (multiple versions of the same data element in one process instance) data errors [17, 18, 19]. A major drawback of these approaches is that they assume that data elements are created and updated exclusively by a single instance of a business process. This assumption does not hold true for interacting business processes with shared data resources. In such cases, creation of new data elements (e.g., new inventory) by one business process may be consumed by a second business process. In this scenario, existing approaches would detect missing and redundant data flow errors for the two business processes. MIRABELLE's extracted workflow model unifies the representation of data and control dependencies, which enables the extensible model-finding formalism to identify a broad range of vulnerabilities.

The principles that apply to showing the output from a scenario generation also apply to showing the output of the identified vulnerabilities. When we find a presumed vulnerability, it is vital to present it in a way that users find comprehensible. However, there are key differences between the prior research on scenario generation and BL flaws. Following programming practice, it is likely that users would want to see minimal reproducible examples (MREs). That is, it must be as small as possible (avoiding superfluous information), reproducible (so they can trace it), and an example (i.e., concrete). Adjusting existing work to handle this case presents interesting technical challenges. For one thing, there is no concrete evidence supporting the value of minimality in MREs; this appears to be folk wisdom, but our prior research shows how folk wisdom on minimality (in other but related settings) can be wrong [20]. Moreover, it is likely that the use of richer cognitive techniques—such as our presentation of negative examples for system understanding—can be valuable here.

Early work converting natural language to business processes combined the Stanford Syntax parser together with other pre-deep learning approaches to form a hierarchical transformation process [14]. A review of pre-deep learning approaches for process mining and extraction can be found in [5]. More recently, LLMs [9] have attracted great interest due to their general-purpose context engineering that allows models to perform novel language tasks at inference time that without back-propagation based retraining. Previous research using LLMs demonstrated the ability to parse BPMN from natural language [6]. In that study, researchers used an iterative model generation approach built from Partially Ordered Workflow Language (POWL) [6] to recursively construct more accurate business process models from simpler sub-models. This approach over comes zero shot inference time generation of XML and JSON that appeared in initial LLM BPMN conversion efforts [8]. More recently, Vision Language Models (VLMs) in combination with OCR have demonstrated the ability to extract BPMN directly from images without requiring text [7]. Despite the remarkable progress converting natural language to BPMN, their zero shot and bespoke pipeline architectures limit their ability to independently iterate towards a satisfactory NL to BPMN translation. In [22] the authors use LLM's to extract high-quality semantics (in the form of triplets) from business process text.

The next section focuses on the document processing aspect of our own MIRABELLE pipeline and the various ways that it employs LLM's in an upstream position in order to complete vital (intermediate) steps which support downstream vulnerability detection.

## 3 Documentation-Processing Toolkit (DPT) with Large Language Models

The MIRABELLE processing system (see Figure 1 for big picture) has several vital processing steps. Figure 2 illustrates the relative position and role of document-processing toolkit (DPT) within that pipeline. In comparison with the state of the art only a few years ago, today's LLMs provide very powerful basis for reading and understanding verbose and descriptive text comprising standards, process guidelines, work instructions, and more. For our experimentation, several pragmatic questions about LLM utilization were:

- Do they reliably comprehend the nuance of common business verbiage?
- Will they show acceptable accuracy and precision when identifying elements?
- Can they effectively extract the logical process structures found in user guide and work instruction documents?

This section focuses mostly on our efforts so far to discover the processing aspects that will benefit most from LLMs and support the end goals of the pipeline – vulnerability and error detection.

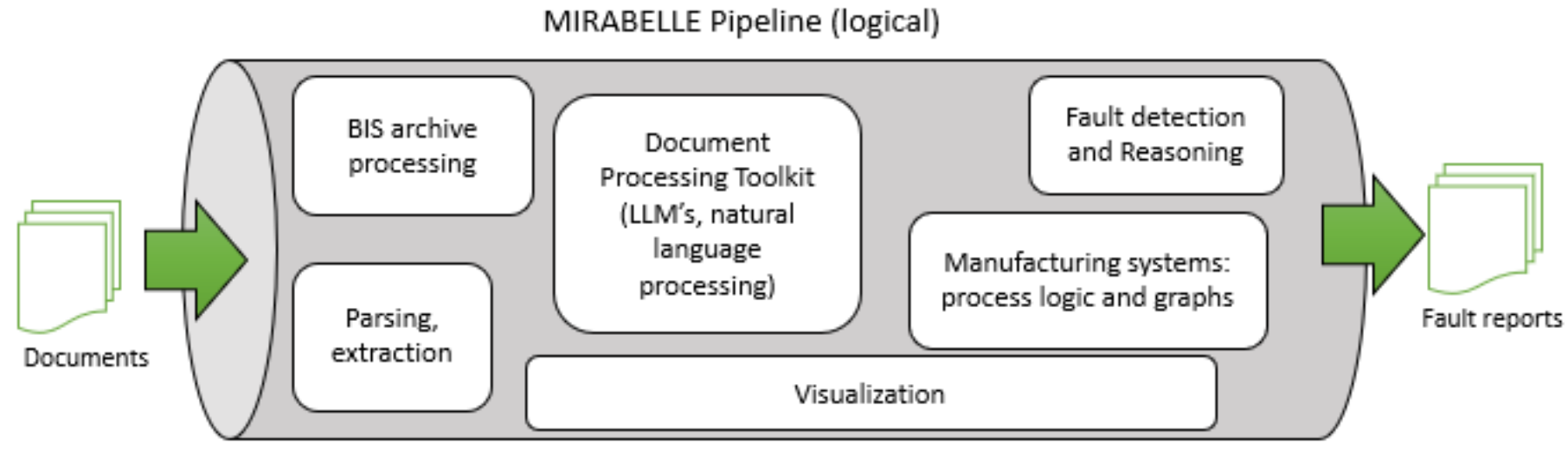

**Fig 2**. Document processing toolkit supports the goal of vulnerability detection.

DPT is a sort of toolkit of APIs and techniques, designed to ingest documentation semi-automatically, and help extract and analyze embedded business logic in support of vulnerability discovery. Broadly speaking, it is built upon a diverse tech stack that includes Python, FastAPI, MLflow, Flask, JavaScript, and Docker. LLMs and supporting NLP tools are served over HTTP from an internal Ollama server. While the controlled (CUI) nature of much of our data required on-premises processing, we have also used cloud/public LLMs in the creation of synthetic datasets for testing prompt and LLM performance. DPT supports the following documentation-processing-related tasks:

1. Ingestion of long-form textual documentation (such as user guides and SOP) and discovery of errors of a contradictory nature; e.g., it *says A here*, and *B somewhere else*, and *A contradicts B*.
2. Ingestion of short work instructions (WI) meant to guide operators through tasks, and the determination of whether contradictions or errors are present. For example, the use of both "exactly" and "at least" when referring to a given action value.
3. Ingestion of long-form textual documentation and reconstruction of the logic described therein, in order to gain insights and to support vulnerability detection.

The next sections describe each of the above efforts and their results.

### 3.1 Contradiction detection with LLM-based classification

We set up experimental tests that ran over business documentation to discover the extent to which LLMs can assist in finding errors related to the consistency of the information. Business documentation can be particularly verbose - standard operating procedures (SOP) define how organizations carry out operations safely, efficiently, and consistently. Standards documents serve as guiding lights and absolute truths. User guides provide detailed instructions for operators. All of this information is vulnerable to costly errors. While documentation is often multimedia in nature, here we focus only on text content.

In this experimentation we leveraged business documentation from collaborators[3] as well as datasets from the Stanford Contradiction Corpora (SCC) [1]. In the SCC a textual contradiction between two passage is present if it would be regarded as *"..very unlikely that both passages could be true at the same time"*. For example, a contradictive pair ("t" and "h" identify the pair parts) of *numeric* class is:

```
{"type":"numeric", "t":"Qeshm Island surface is mostly rocky and barren with a small human population of approximately 8,000.", "h":"The population of Qeshm Island is about 200,000."}
```

Contradictive text in a manufacturing SOP's, for example, might manifest as verbiage such as, "*..tighten the bolt B3 using the wrench to no more than 45 ft-lbs*" in one place, and then the inclusion verbiage such as, "*..recall that bolt B3 should be tightened to at most 90 ft-lbs at all times*", in another[4]. In our experimentation we used elements in the real-life SCC to play the role of the kind of documentation errors that are commonly introduced by careless human authors. They also served as a proving ground for the LLM's general ability to discover such faults. The main hypothesis of this experiment is that LLMs can detect these lexical contradictive passages at accuracy of .75 or more when they are embedded within in a large documentation set. The method of the experiment can be described as follows:

- Use an exemplary document(s) as a host document (HD); ideally, it should be both typical and lengthy (note that docs may be pre-processed into a single HD);

[3] Proprietary SOP-type documentation from our collaborators cannot be shared here
[4] The over-cranking of even a single bolt can cause catastrophic damage in sensitive settings

- Chunk HD into M chunks (pages or sections may be “natural” divisions);
- Choose N contradictions from the SCC (by class, by mix, etc.) and inject them into the HD by choosing different random chunks for the “t” and “h” parts of the contradiction. Store all injection metadata;
- Pass all chunk-pairs to the LLM, prompting it to recover inconsistencies;
- Determine accuracy of the run based on ground truth.

As for experimentation setup, we used the following LLMs hosted on on-premises inference servers[5]: Smoll2:1.7B, Gemma3:27B, Llama3.3:70B and Q5KM, and GPT-OSS:120B. These models were not fine-tuned for the experiment. On each run, we held the following variables steady: LLM model, chunk size, chunk overlap (e.g., 128-tokens was common), error injection count, and the LLM hyperparameters *num_ctx* and *temperature*. Aspects of the LLM prompting are shown below.

```
Consider the two text passages tagged by |section1| and |section2|.  If there
is any inconsistency or factual contradiction between the two passages, then
describe it in one or two sentences without extra commentary. However, simply
being unrelated does not constitute a reportable error. <omitted>.
```

**Scaling via LLM-as-Classifier.** In order to handle the quadratic explosion in the number of chunk pairings in this experiment we designed an approach that would strive to generate as few tokens as necessary while evaluating chunk pairs for contradiction. To do this, we used the *logprob* outputs of the LLM on the single token responses (“yes”, “no”) to both positive-sense and negative-sense questions about pair contradictions [23]. From these we could estimate the probability of contradiction for any pair, and we only ran deeper dives into pairings whose contradiction probability exceeded a threshold that we set. As a pragmatic matter, we preferred to run without using this technique and we strove to contain the size of the host document and the number of pairs so as they would complete in reasonable times on our lab hardware[6].

**Results.** Figure 3 illustrates some of the results of this experimentation. We found the best results with GPT-OSS:120B, with adequate accuracy using a chunk size of 512 tokens. Interestingly, the LLM accuracy showed a positive trend as the chunk size became smaller (e.g., 512 token size was better than 4096, 2048, and 1024 tokens). We also found a general increase in accuracy as the size of the chosen LLM model changed from SmolLM2 (1.7B parameters), to Gemma3 (27B), Llama3.3 Q4 (70B), Llama3.3 Q5KM (70B) and finally to GPT-OSS (250B parameters). SmolLM2 failed dramatically, with accuracies near 0, while Llama3.3Q4 and Q5KM showed quite poor accuracy in the .4-.5 range.

---

[5] Hardware: AMD Ryzen Threadripper 7000 Pro, NVIDIA RTX 6000 Ada w. 48GB VRAM

[6] The technique is often referred to as “LLM-as-classifier” (or predictor) and it can be an effective scaling technique, allowing the early culling of pairs not likely to exhibit contradictions, while generating only 1 or 2 output tokens [23]; however, not all LLM’s provide the necessary statistics in their outputs.

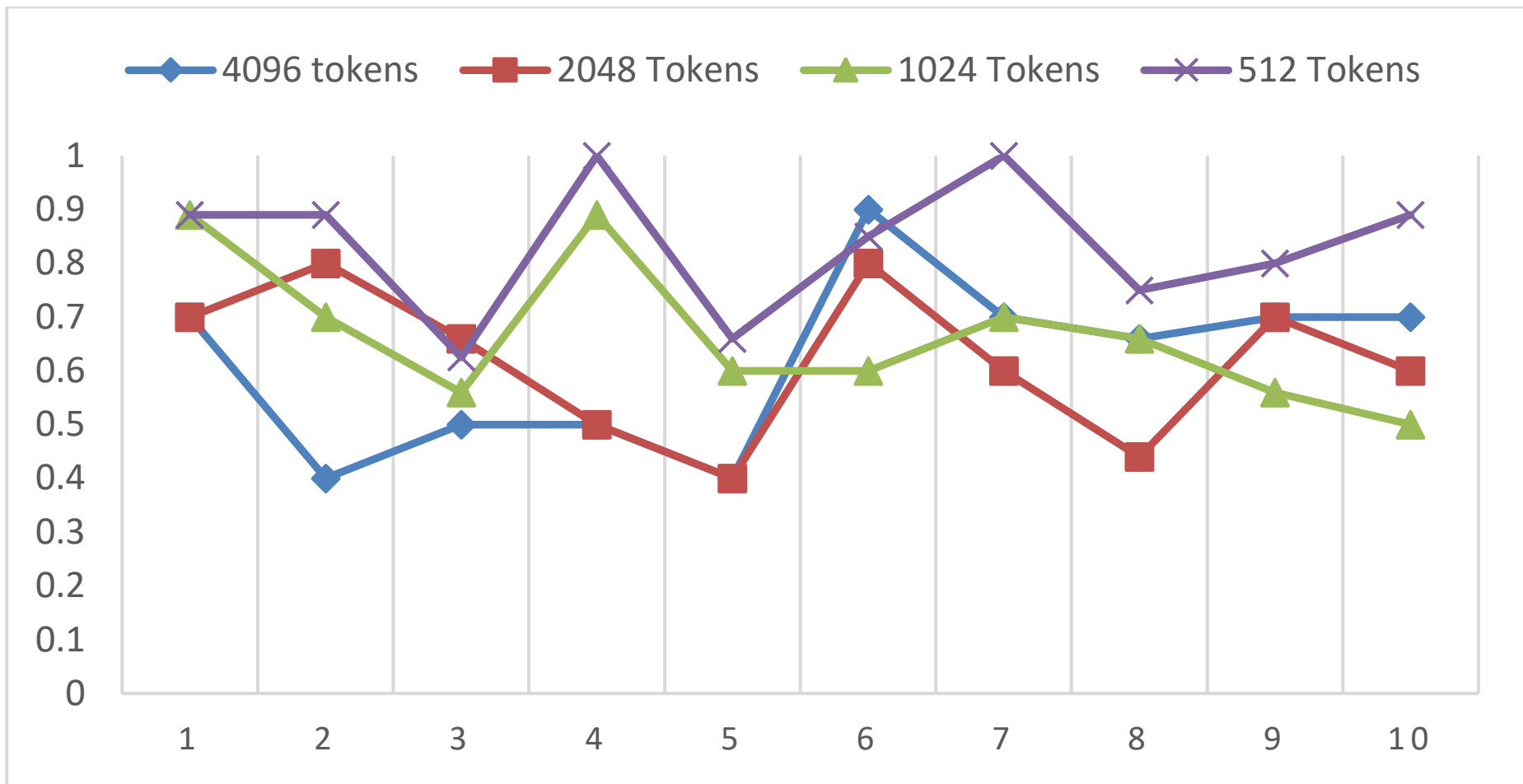

**Fig 3**. GPT-OSS:120B accuracy in the contradiction detection task across document chunk sizing series.

The hypothesis that LLMs can be effective for uncovering documentation contradictions buried deeply in business text is shown to be valid. Several factors had profound effects on accuracy, most notably the LLM model and the chunk size, and the following caveats should be noted. First, the text payloads of SCC contradictions are not pure business text and this lack of contextuality might give the LLM an unfair advantage in finding them in the first place. Secondly, the number of chunk pairs P grows in polynomial size in the number of chunks M. For example, in a given run, if M=50 chunks, then more than 1,225 chunk pairs are separately verified by the LLM. This makes parallelization and machine learning operations quite important in order to scale to very large host documents.

### 3.2 Flagging erroneous Work Instructions

So-called work instructions are essential text passages (sometimes including linked images) that are presented to users at each step of manufacturing work system operations. They are vital aspects of the workflow, and not immune to errors. We set up experimental LLM flows to determine the extent to which LLM prompting could uncover subtle work instruction errors. An exemplary textual work instruction is:

```
Install the Cooling Fan. Mount Fan to chassis. Torque the bolts to not more than
20 ft/lb and record results with the Digital Torque Wrench.
```

It is not difficult to imagine how small errors in these instructions, such as typos, units of measure mismatches, or conversions, could create a crisis of cascading faults on an assembly line. Our intuition told us that even pure natural language processing (NLP) would struggle with, for example, unit conversion errors within instructions. At least, without additional external supporting tools. Also, quantity qualifiers such as "at least", "at most", "not more than" are often found within the instruction in non-trivial combinations. For this experimentation setup and dataset, we did not have access to a large volume of erroneous work instructions, so we provided GPT-OSS:250B with an example set and used it to generate a set of 100 high-quality synthetic work instructions, half with subtle errors and half without. The sets were checked by hand for quality. The following is an exemplary erroneous instruction in which the converted pressure limit is incorrect (750kPa does *not* equal 5bar) and the "no higher than" qualifier contradicts with the "exactly" qualifier:

```
Pressurize the fuel rail to exactly 5bar (750kPa) and verify the gauge reads no
higher than 4.5bar.
```

Our hypothesis was that an LLM could flag a satisfactory number of these kinds of work instructions with both accuracy and precision. Our setup included the dataset mentioned above, and the GPT-OSS:250B model set with temperature setting of 0.1. Prompting technique was akin to the following:

```
 [...] if there is any obvious or clear inconsistency or factual contradiction
within the text then reply with a brief explanation of the issue. If there is
no such issue, then respond only with 'none'. Do not make irrelevant or extra-
neous commentary.
```

First, we ran the LLM on the clean, error-free instructions set to test precision. Here, the LLM exceeded expectations, correctly validating all the clean instructions. It found a zero-day error in a problematic "clean" instruction that we missed; the instruction, in fact, had ambiguous text. Thus, precision was 1.0 (or 0.96 if the let the zero-day error ride). The next task involved running the LLM over the error set to test accuracy. The LLM again scored very highly with a score of 0.98. In fact, it missed only 1 of the 50 errors scoring the slightly controversial (erroneous) work instruction below as "clean":

```
Mount the oil pump gear using a preload of 0.5mm and verify the clearance is
less than 0.5mm.
```

As we looked again at the above instruction missed by the LLM we noticed that it was not completely obvious that "clearance" and "preload" refer to the same artifact. At any rate, even with this "miss", the LLM accuracy was very high.

**Results.** As shown in Table 1, our hypotheses that the LLM would have both high precision and accuracy in flagging erroneous work instructions containing conversion errors, range-limiting mistakes, and numerical mistakes, was corroborated.

**Table 1.** Summary of LLM accuracy and precision when run over a set of 100 erroneous work instruction corpus

| Metric | Score | Comments |
|---|---|---|
| Precision | 0.96 | LLM did not find errors in instructions where they "were not" |
| Accuracy | 0.96 | LLM correctly found all faulty work instructions (WI), often identifying all problems in a WI, even if multiple |

### 3.3 Automated LLM-Driven Extraction, Modeling, and Petri Net Generation from ERP User Guides

Enterprise Resource Planning (ERP) systems rely on detailed technical documentation to describe operational procedures, user roles, and interdependent workflows. ERP guides exemplify this complexity, containing multi-step processes, conditional logic, and extensive cross-references that are difficult to formalize manually. Traditional workflow extraction methods are slow, error-prone, and require significant domain expertise. To address these challenges, we developed an automated pipeline that leverages Large Language Models (LLMs) to transform ERP user guides into structured workflow models and executable Petri Nets. The system performs complete end-to-end processing—PDF ingestion, structure discovery, reference extraction, user action summarization, workflow decomposition, and CPN-Py JSON generation—without human intervention. These machine-readable outputs support simulation, verification, auditing, and integration with enterprise analysis tools. This section describes the pipeline architecture, processing stages, and comparative results across three LLM backends.

**System Architecture**. The architecture supports both Controlled Unclassified Information (CUI) and non-CUI environments without altering pipeline behavior. Users may select from three backend configurations: OpenAI's non-CUI commercial LLMs, a locally hosted LLM on an RTX 6000 GPU for secure on-premise inference, and Azure-hosted Llama 3 on A100 GPUs for scalable CUI processing. Despite differences in trust boundaries and compute capabilities, the pipeline maintains uniform logic and identical artifacts across environments.

A standardized Python stack orchestrates PDF parsing, markdown normalization, hierarchical chunking, and LLM-driven extraction. PyMuPDF4LLM captures headings, page text, images, and structural metadata; mdformat ensures consistent markdown; and LangChain manages chunking and prompt orchestration. Containerization ensures reproducibility across developer, cloud, and secure enclave deployments.

An LLM Ops layer, built on MLflow, provides versioned prompts, controlled experimentation, and systematic evaluation of model outputs. This governance framework enables reproducibility, cross-model comparison, and traceability of changes in both development and production environments.

**End-to-End Processing Pipeline**. The automated pipeline consists of four stages. **Stage A** (PDF Parsing and Markdown Conversion) extracts text, headings, section hi-erarchy, and embedded images, producing a normalized markdown representation that preserves document structure. **Stage B** (Hierarchical Recursive Chunking) splits the markdown into semantically coherent chapter- and section-level units. This structure-aware decomposition prevents context mixing in LLM prompts and improves both accuracy and processing efficiency. **Stage C** (LLM-Based Information Extraction) identifies external references to other Oracle guides, internal cross-references within the document, and concise user-action summaries. These summaries condense long procedural text into structured action sequences. **Stage D** (Workflow Extraction and Petri Net Generation) translates the extracted actions and dependencies into CPN-Py-compatible JSON. The resulting workflows support Petri Net visualization, simulation, and automated analysis. Across multiple Oracle guides, this pipeline has executed successfully without manual data transformation.

**Case Study: Oracle Purchasing User Guide**. To evaluate the system, the pipeline was applied to the Oracle Purchasing User Guide, with special attention to Chapter 15 (Receiving), a chapter known for complex procedures and dense cross-referencing. The guide was processed fully automatically, demonstrating the system's ability to preserve structural fidelity and accurately extract procedural information from real-world ERP documentation. Three LLM backends were evaluated: OpenAI GPT-5.2 (non-quantized), Azure-hosted quantized Llama 3, and locally deployed quantized Gemma 3. These configurations represent a spectrum of reasoning depth, quantization effects, and throughput characteristics, enabling comparative assessment across heterogeneous execution environments. The processing pipeline here has several phases, and the prompting that extracts business process semantics from the marked-down User Guide text is structured into detailed descriptions of role, task, format constraints, schema and output requirements.

**Results**. Across all models, extraction tasks—external reference detection, intertextual mapping, and action summarization—were completed with high fidelity. See Table 2 for details. The similarity of extraction outputs indicates that markdown normalization and hierarchical chunking effectively preserve context for LLM processing, allowing quantized models to approach the performance of larger non-quantized models.

**Table 2.** Processing times for Chapter 15 (Receiving) of the Oracle Purchasing User Guide

| Step | OpenAI API Platform GPT 5.2 | Azure Cloud Ollama Llama 3 (Quantized) GPU: A100 | On Prem Ubuntu Ollama Gemma 3 (Quantized) GPU: RTX 6000 |
|---|---|---|---|
| A. PDF Extraction | 2 Mins 40 Secs | | |
| B. Markdown Chunking | 2 Secs | | |
| C1. Extract References | 50 Secs | 27 Secs | 2Mins 28 Secs |
| C2. Summarize References | 43 Secs | 15 Secs | 37 Secs |
| D. Business Process | 2 Mins | 50 Secs | 3 Mins 41 Secs |
| **Total (Chapter 15)** | 5 Mins 15 Secs | 4 Mins 14 Secs | 9 Mins 28 Secs |

Significant differences emerge in structured workflow generation. GPT-5.2 produced the most accurate and coherent CPN-Py JSON, consistently capturing nuanced procedural dependencies. In contrast, quantized Llama 3 and Gemma 3 generated workflows that were usable but occasionally incomplete or structurally inconsistent, especially for deeply nested or highly interdependent processes.

These discrepancies reflect the reduced reasoning capacity associated with quantization. Benchmarking revealed a clear trade-off: quantized models significantly outperform GPT-5.2 in processing speed for extraction-oriented tasks, but GPT-5.2 provides superior accuracy in multi-step reasoning and schema-aligned Petri Net generation. Qualitative comparisons confirmed that reference extraction is stable across models, while structured workflow synthesis remains sensitive to model capacity. Ongoing work aims to mitigate these limitations through multi-prompt and multi-agent strategies that distribute complex reasoning tasks.

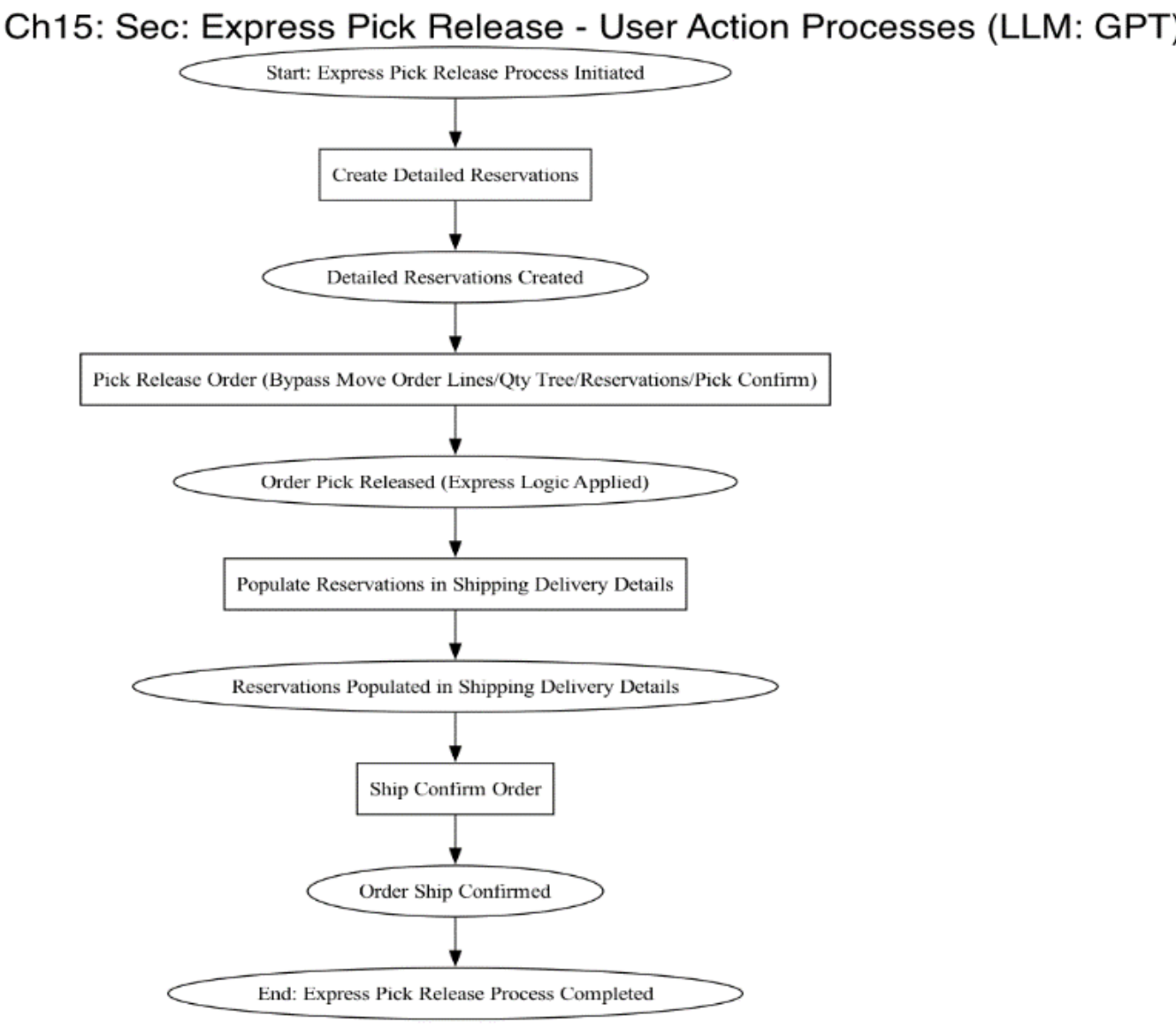

**Fig 4**. Resulting Petri Net representation of one section of the User Guide

Oracle's Purchasing User Guide[7] is the source material for at least one of the experiments here. The content of its pages varies from sparse textual pages with figures depicting user interfaces to those that are text-only with roughly 500 words per page. Intra-document linkages within this very voluminous information set sometimes interconnect the chapters and sections, creating interdependent processes. For example, verbiage similar to, "see Section 14 for shipping location definitions" creates implicit connectivity.

[7] https://docs.oracle.com/cd/E26401_01/index.htm

Figure 4 is an exemplary output from the processing, using GPT-5.2, wherein a Petri Net structure is extracted from a particular section of the User Guide relating to Pick Release steps, typically a bridge between the sales offices and the warehouse floors. In the Petri Net figure, ovals are places (e.g., storage, buffers, etc.) and rectangles are transitions.

**Discussion**. Business Process Model and Notation Operationally, containerized environments and MLflow-based governance support reproducibility, cross-model comparisons, and stable deployment across CUI and non-CUI settings. Broader insights include: the critical importance of preserving document hierarchy; the speed advantages but reasoning limitations of quantized models; and the superior structured-generation performance of non-quantized GPT-level models.

### 3.4 Summary of Document Processing Toolkit

This section has described several techniques in out document-processing toolkit in which LLMs are pointed at enterprise information for the purposes of vulnerability detection. As discussed, our lab hardware setup allows for moderately large LLMs to be utilized, and we have satisfactory results on our tasks from both large and smaller LLMs. DPT is currently: 1) finding tricky conversion and grammatical errors in work instructions, 2) finding grammatically contradictive text within large documentations, and 3) recovering the structure of business processes (as graphs) from large inter-related documentation sets.

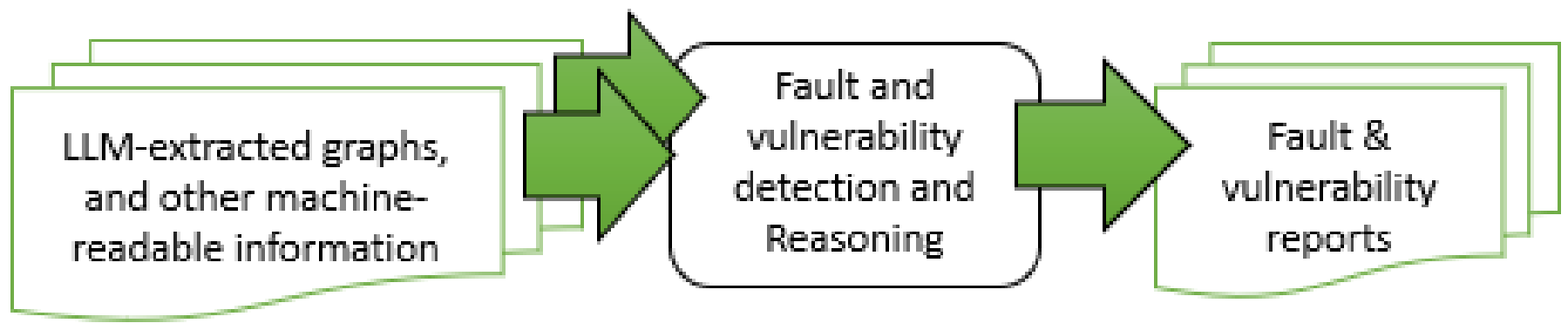

**Fig 6**. Completing the pipeline from LLM-extracted insights to vulnerability discovery

All of the above moves us closer to automated LLM-assisted vulnerability discovery. While the details are out of scope in this paper, we remark that results from DPT components are 1) encoded in machine-readable schemata, 2) optionally converted into alternate representations, and 3) propagated to classical reasoning components further downstream. Our reasoning components create the fault and vulnerability report (see Figure 6). In the next section we conclude and describe future work.

## 4 Conclusions and Future Work

MIRABELLE demonstrates that leveraging LLMs significantly accelerates the transition from unstructured enterprise documentation to rigorous, machine-readable formats. Our experimentations show that only moderately complex prompting techniques effectively extracts business logic contradictions across various model scales, while achieving high accuracy in detecting nuanced conversion and range errors within work instructions. In addition, the successful generation of Hierarchical Colored Petri Nets (HCPN) from online documentation bridges the gap between natural language and classical logic fault detection, and we have developed a nearly fully automated process for this phase. These achievements validate our ongoing work combining the linguistic flexibility of generative AI with the precision of formal logic to identify critical industrial process faults.

Our current and future work includes further refinement of prompting strategy to improve accuracy of graph extraction, direct integration of DPT with the classic reasoning components, and further testing with industrial data from partners.

**Acknowledgments.** The authors acknowledge the support of the research team. LLM's were used while composing this paper for the purposes of light edits and grammar-checking.

**Disclosure of Interests.** This material is based upon work supported by the NIWC Atlantic under Contract No. N6523624C8007

## References

1. Marneffe M., Rafferty A., Manning C.: "Finding contradictions in text". *46th Annual Meeting of the Assoc. for Computational Linguistics and Human Language Technology Conference*, Ohio (2008).
2. Kourani H. et al. "ProMoAI: process modeling with generative AI." *arXiv preprint arXiv:2403.04327* (2024).
3. Katsios G. et al. "MetaBPL: Fault Detection in Business Logic Systems." *Human Aspects of Advanced Manufacturing, Production Management and Process Control* 184 (2025): 102.
4. Friedrich F., Mendling J, and Puhlmann F. "Process model generation from natural language text." *International conference on advanced information systems engineering*. Berlin, Heidelberg: Springer Berlin Heidelberg, 2011.
5. Bellan, P., Dragoni M., and Ghidini C. "A qualitative analysis of the state of the art in process extraction from text." *Proceedings of the AIxIA 2020 Discussion Papers Workshop*. Vol. 2776. CEUR-WS. org, 2020.
6. Kourani H., and J. van Zelst. S., "POWL: partially ordered workflow language." *International Conf. on Business Process Management*. Cham: Springer Nature Switzerland, 2023.
7. Deka P., and Devereux B. "Structured Extraction from Business Process Diagrams Using Vision-Language Models." arXiv preprint arXiv:2511.22448 (2025).
8. Berti, A., Schuster D., and MP van der Aalst, W. "Abstractions, scenarios, and prompt definitions for process mining with llms: A case study." *International conference on business process management*. Cham: Springer Nature Switzerland, 2023.
9. Achiam, J. et al. "Gpt-4 technical report." *arXiv preprint arXiv:2303.08774* (2023).
10. Celonis Inc., "Process Mining Tools," [Online]. Available: https://celonis.com.
11. SAP, "SAP Signavio," [Online]. Available: https://www.signavio.com/.
12. ProM Tools. "Process MIning Workbench." https://promtools.org/prom-6-12/.
13. Camunda, "Zeebe YAML workflows", [Online]. Available: https://unsupported.docs.camunda.io/0.25/docs/components/zeebe/yaml-workflows/.

14. Friedrich F., Mendling J., and Puhlmann F. “Process Model Generation from Natural Language Text.” *In Int’l. Conference on Advanced Information Systems Engineering*, 2011.
15. Honkisz K., Kluza K., and Wiśniewski P., “A Concept for Generating Business Process Models from Natural Language Description.” *In 11th International Conference on Knowledge Science, Engineering and Management*, KSEM 2018, Changchun, China, August 17–19, 2018.
16. Sonbol R., Rebdawi G., and Ghneim N. “A Machine Translation Like Approach to Generate Business Process Model from Textual Description.*”, SN Computer Science*, 4(3), Mar 2023, https://doi.org/10.1007/s42979-023-01742-z.
17. Stackelberg S.V., Putze S., Mülle J., Böhm K., “Detecting Data-Flow Errors in BPMN 2.0.” *Open Journal of Information Systems*, 1(2), (2014): 1-19.
18. Calvanese D. et al, “Formal Modeling and SMT-Based Parameterized Verification of Data-Aware BPMN.” Available at: https://doi.org/10.48550/arXiv.1906.07811.
19. Liu C. et al., "Petri Net Based Data-Flow Error Detection and Correction Strategy for Business Processes", IEEE Access, vol. 8, pp. 43265 - 43276, 2020, doi: 10.1109 / ACCESS.2020.2976124.
20. Dyer T., Nelson T., Fisler K., and Krishnamurthi S., “Applying Cognitive Principles to Model-Finding Output: The Positive Value of Negative Information.” *ACM SIGPLAN conference on Object-Oriented Programming Systems, Languages & Applications*, 2022.
21. Object Management Group, "Business Process Model Notation 2.0," [Online]. Available: http://www.omg.org/spec/BPMN/2.0/.
22. Zhang, B., & Soh, H. “Extract, define, canonicalize: An LLM-based framework for knowledge graph construction”. *Proc. Empirical Methods in Natural Language Processing* (2024).
23. Brown, T. B., Mann, B., Ryder, N., Subbiah, M., Kaplan, J. D., Dhariwal, P., Neelakantan, A., Shyam, P., Sastry, G., Askell, A., et al., “Language models are few-shot learners”. *Advances in Neural Information Processing Systems*, *33*, 1877–1901, 2020.